# Transient Attack against the KLJN Secure Key Exchanger

Shahriar Ferdous [+,1] and Laszlo B. Kish [1,2]

[1] *Department of Electrical and Computer Engineering, Texas A&M University, College Station, TX 77843-3128, USA*

[2] *Óbuda University, Budapest, Bécsi út 96/B, Budapest, H-1034, Hungary*

[+]Corresponding author: ferdous.shahriar@tamu.edu

## ABSTRACT

We demonstrate the security vulnerability of the ideal Kirchhoff-Law-Johnson-Noise (KLJN) key exchanger against transient attacks. Transients start when Alice and Bob connect the wire to their chosen resistor at the beginning of each clock cycle. A transient attack takes place during a short duration of time, before the transients reflected from the end of Alice and Bob mix together. The information leak arises from the fact that Eve (the eavesdropper) monitors the cable, and analyzes the transients during this time period. We will demonstrate such a transient attack, and, then we introduce a defense protocol to protect against the attack. Computer simulations demonstrate that after applying the defense method the information leak becomes negligible.

## INTRODUCTION

A secure key exchanger is an integral part of secure communications. Information-theoretic[1-7] (unconditional) secure key exchange is offered by Quantum Key Distribution (QKD)[8-44] and its statistical-physical alternative, the Kirchhoff-Law-Johnson-Noise (KLJN) secure key exchanger[3-7,45-99]. QKD utilizes the quantum no-cloning theorem; as opposed to KLJN scheme, which is based on classical statistical physics, particularly the Fluctuation Dissipation Theorem[3], Gaussian stochastic processes[63] and thermal equilibrium[97].

Even though the ideal KLJN scheme offers perfect security[1-7], the practical systems have non-ideal features that Eve can exploit for passive (passively listening) attacks. The KLJN system has been subject to various passive attacks, including transient attack[45-49].

The present paper demonstrates such an attack and proposes a defense method against it. It is shown by computer simulations that new defense mechanism successfully eliminates the information leak, thus it nullifies the attack.

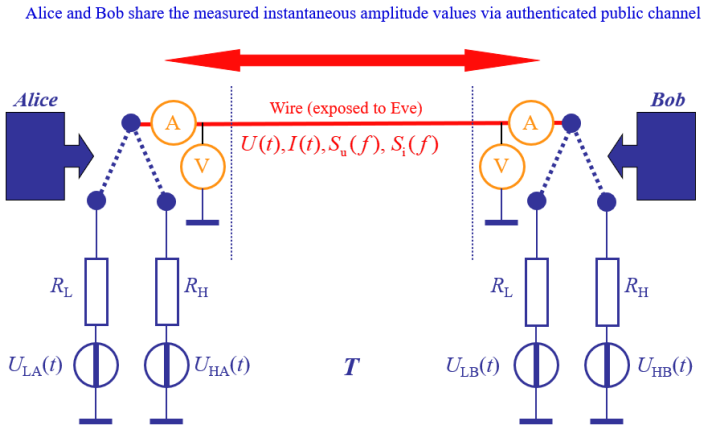

**FIG. 1.** The core of the KLJN secure key exchanger scheme consists of a wire line connection between the two communicating parties Alice and Bob[50-51]. The voltage $U(t)$, the current $I(t)$, and their spectra $S_u(f)$ and $S_i(f)$, respectively, are measurable by Alice, Bob and Eve. In the private space of Alice and Bob, the voltage generators $U_{LA}(t)$, $U_{HA}(t)$, $U_{LB}(t)$ and $U_{HB}(t)$ represent the independent thermal (Johnson-Nyquist) noises of the resistors, or optionally they are external Gaussian noise generators for higher noise temperature. The homogeneous temperature $T$ in the system guarantees that the LH (Alice $R_L$, Bob $R_H$) and HL (Alice $R_H$, Bob $R_L$) resistor connections provide identical mean-square voltage and current, and the related spectra, in the wire[3,5,50-51], and that the net power flow is zero between Alice and Bob.

**The KLJN key exchanger** [50-51]

The core of the Kirchhoff-Law-Johnson-Noise (KLJN) system[3-7] is shown in Fig. 1. It consists of a wire line connecting the communicators at Alice's and Bob's sides. The temperature is homogeneous in the whole system (thermal equilibrium situation) and, in each communicator, there are two switches and a resistor pair, $R_H$ and $R_L$, (where $R_H > R_L$ and $R_H \neq R_L$). The pairs are identical at Alice's and Bob's sides[3,5,50-51]. At the start of a secure *bit exchange period* (BEP), each party (Alice or Bob) randomly chooses one of the resistors ($R_H$ or $R_L$) and connects it to the wire line for the whole BEP. This process yields a new securely shared bit[3,5], see below. The HL ($R_H$ at Alice and $R_L$ at Bob), and the LH ($R_L$ at Alice and $R_H$ at Bob) situations represent the two states of the possible values of the shared secure bit, respectively. Alice and Bob publicly agree about the interpretation of the bit values (0 or 1) for the HL and the LH situations, respectively[50-51].

The noise spectra $S_u(f)$ and $S_i(f)$ of the voltage $U(t)$ and current $I(t)$ in the wire, respectively, are given by the Johnson-Nyquist formulas of thermal noise[3,5]:

$$S_u(f) = 4kTR_p, \quad (1)$$

$$S_i(f) = \frac{4kT}{R_s}, \quad (2)$$

where $k$ is the Boltzmann constant, $T$ is the noise temperature and $R_p$ and $R_s$ are the parallel and serial resultant values of the connected resistors, respectively. In the LH and HL cases, the resultant values are:

$$R_{pLH} = R_{pHL} = \frac{R_L R_H}{R_L + R_H}, \quad (3)$$

$$R_{sLH} = R_{sHL} = R_L + R_H. \quad (4)$$

The quantities that Eve can access with passive measurements satisfy the following equations that, together with Equations (3)-(4), form the pillars of security regarding passive attacks against the KLJN scheme[50-51]:

$$U_{LH} = U_{HL}, \quad (5)$$

$$I_{LH} = I_{HL}, \quad (6)$$

$$P_{LH} = P_{HL} = 0, \quad (7)$$

where the voltage $U$ and current $I$ values stand for the effective (RMS) amplitudes in the wire, $P$ is the mean power flow between Alice and Bob, and the LH and HL indexes stand for the secure bit values represented by connected resistances, see above.

Equations (5) and (6) are the consequences of Equations (1)-(4), implying that the LH and HL resistor situations provide the same mean-square noise voltage spectra and noise current spectra, because both the parallel and serial equivalent resistances are identical. Equation (7) is due to the thermal equilibrium situation[3,5,50]. With passive eavesdropping (by measuring the wire voltage and current), Eve can determine the resultant (both parallel and serial) values of the connected resistances by evaluating the noise voltage and current spectra[50-51]. However she is unable to differentiate between the LH and HL situations, see below.

If, during the BEP, both parties connect to the same resistance value HH ($R_H$, $R_H$) or LL ($R_L$, $R_L$), the situation is not secure, see Fig. 2, because then Eve can determine the resistor value at the other side[50-51]. Thus, the HH and LL bit situations are discarded. The only secure combinations are the HL ($R_H$, $R_L$) and LH ($R_L$, $R_H$) cases[3,5]; because determining the resultant resistance in the loop, does not provide Eve with enough information to break the code[3,5]. In the HL or LH cases, Eve cannot determine which side has $R_H$ and which side has $R_L$, thus she does not know if the state is HL or LH, which means she does not know if the key bit value is 0 or 1. For Eve this is an information entropy of 1 bit indicating perfect unconditional security[50-51].

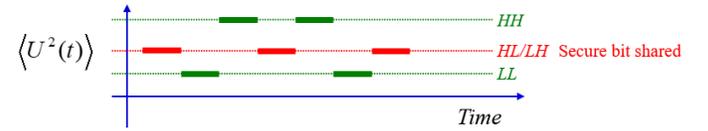

**FIG. 2.** Illustration of what Eve sees regarding the voltage: the mean-square voltage levels in the wire. The H and L indexes stand for the chosen resistors. The HL and LH levels are identical which makes the bit values corresponding to the HL and LH cases secure. A similar graph exists for the mean-square currents[50-51].

## TRANSIENT ATTACKS

Transients are propagating voltage and current fronts in the cable. They start when Alice and Bob connect the wire to their chosen resistor at the beginning of each clock cycle and voltage and current suddenly change. The transients and reflections depend on the terminating resistance of the cable thus they can be utilized for attacks.

A transient attack[45-49] takes place during a short duration of time before the voltage and current in the cable reach thermal equilibrium. The information leak arises from the fact that Eve can monitor the cable and analyze the transients during this time period.

Let's assume, Alice connects to $R_H$ and Bob connects to $R_L$; which is the HL situation. It takes a single fly time $t_f = \frac{L}{c}$, where $L$ is the length of the cable and $c$ is the propagation velocity of EM waves in the cable, for the transient to propagate from one side to the other. During the transient phase, Eve can monitor, for example, the evolution of the square of voltage at both ends and can distinguish between the HL and LH states[47]; hence cracking the KLJN scheme.

As soon as the noises in the wire begin to mix which later leads to thermal equilibrium[45-49]; transient attacks will no longer pose a threat.

**Former papers on transient attacks against the KLJN scheme**

Transient attack was first officially mentioned in Ref 45; their proposed defense was to linearly ramp up the noise levels by enforcing the envelope of the time functions of noise voltage and current to zero before the switching[45].

Later a new defense was proposed against the transient attack[46]. In order to execute the key exchange, they used continuously variable resistors (such as potentiometers, etc.). At the beginning of the KLJN clock period, both Alice and Bob start with equal resistance on each size, equaling $\frac{R_H + R_L}{2}$. Then Alice and Bob execute independent, slow, continuous-time random-walks with their resistor values (where the system is changing in the adiabatic limit)[46]. The independent random walk takes places within a publicly pre-agreed time period ($t_r$). If within this time period Alice and Bob reach their randomly preselected $R_H$ or $R_L$ value, they stop the random walk and stay at this value. Then, after the $t_r$ time, they start the KLJN protocol, in the regular way[46].

GAA re-introduced[47] the issue of a transient attack. First, they proposed the same defense as it is described above. Then they came up with another defense method: "...*modifying the temperature profile according to the choice of resistor such that the injected signal is initially identical, irrespective of the resistance chosen*."[47]. However, we will show below in the Demonstration section of the present paper that this and similar defense methods are flawed due to several reasons, including the cable reflections that are different at the two ends.

In the present paper, we point out that, for an efficient defense, the noises at the two ends must be starting from zero; and their initial slopes must be properly selected.

**THE PROPOSED DEFENSE METHODS**

Below, a new defense method against transient attacks is introduced. It has two key features:

i. The instantaneous value of the noise voltages starts at zero.

ii. Properly matching the starting slope ratio $m_{HL}$: The starting slope of the noise samples are maintained at a ratio:

$$m_{HL} = \frac{m_H}{m_L} = \frac{R_H + Z_0}{R_L + Z_0}, \quad (8)$$

where, $Z_0$ is the cable characteristic impedance, $m_H$ and $m_L$ refer to the starting slope of the two noise generators at Alice and Bob, respectively, and the H and L indexes refer to the $R_H$ and $R_L$ terminating resistors, respectively.

Feature 1 prohibits step-function type initialization and related reflections from the other end of the cable (see Fig. 3 below). The reflection coefficients $\Gamma_H$ and $\Gamma_L$ at termination by $R_H$ and $R_L$, respectively are:

$$\Gamma_H = \frac{R_H - Z_0}{R_H + Z_0}, \quad (9)$$

$$\Gamma_L = \frac{R_L - Z_0}{R_L + Z_0}, \quad (10)$$

which give information about the connected resistances when the two initial voltage values are identical and non-zero, see in the Demonstration section.

Feature 2 will allow the cable voltages and currents at the ends of Alice and Bob to ramp up similarly during the fly time of the cable, before the noise mixing happens, see also in the Demonstration section.

To provide features 1 and 2, Alice and Bob must *pre-generate and store* their own long and independent noise records. They publicly agree about the common starting slope of the voltages for $R_H$ and $R_L$. Then they perform a search over their own records to find a zero-crossing with a reasonably well-matching slope after it.

**DEMONSTRATION OF THE TRANSIENT ATTACK AND DEFENSE VIA COMPUTER SIMULATIONS:**

For the generation of random Gaussian noise with satisfactory properties, we used MATLAB, and for the transient simulations, the industrial cable simulator LTspice was utilized.

We used a 6-time oversampling with the same factor of zero-padding to prohibit aliasing errors, and to generate a smooth, Gaussian, band-limited white noise. The correlation time of the noise was 10 times longer than the fly time through cable.

The HL state was arranged for all the simulations, where the resistance at Alice's and Bob's ends were $R_H$ and $R_L$, respectively. The resistance values were $R_H = 11$ KΩ, $R_L = 2$ KΩ; the cable characteristic impedance was $Z_0 = 50$ Ω; the noise bandwidth was $B = 5$ KHz; and the noise temperature was $T = 7 \times 10^{15}$ K. The cable length was $L = 2000$ m, which yielded a cable fly time $t_f = 10^{-5}$ s.

During the determination of the zero crossing an ~0.1% margin of error was tolerated. For finding matching slopes ~1% margin of error was allowed compared to the publicly agreed slope values that satisfied Equation (8).

Via LTspice, the mean-square noise voltages and currents were calculated at Alice's and Bob's end, up to the simulation run time, imitating Eve's transient measurement and the related attack. Eve's protocol was to identify the side with the greater and lower mean-square voltages (or currents) as terminated by $R_H$ and $R_L$, respectively.

Mathematically formulating, the above attack protocol is as follows. Assume the mean-square voltages during the transient are $\langle U_a^2(t) \rangle_\tau$ and $\langle U_b^2(t) \rangle_\tau$ at Alice and Bob, respectively. Note, the averaging takes place during the transient observation time $\tau$, which is much shorter than the whole BEP. According to Eve's protocol outlined above, the parameter $\rho_u(\tau)$ is evaluated:

$$\rho_u(\tau) = \langle U_a^2(t) \rangle_\tau - \langle U_b^2(t) \rangle_\tau, \quad (11)$$

where $\rho_u(\tau) > 0$ and $\rho_u(\tau) < 0$ imply that Eve guesses HL or LH for the particular situation, respectively.

Similar equation can be derived for the mean-square noise current, see Equation (12).

$$\rho_i(\tau) = \langle I_a^2(t) \rangle_\tau - \langle I_b^2(t) \rangle_\tau, \quad (12)$$

where the mean-square noise current is $\langle I_a^2(t) \rangle_\tau$ and $\langle I_b^2(t) \rangle_\tau$ at Alice and Bob, respectively. And, $\rho_i(\tau) > 0$ and $\rho_i(\tau) < 0$ imply that Eve guesses HL or LH for the particular situation, respectively.

Below we demonstrate via computer simulation that there will be information leak due to transient attack, if there is no defense. The duration of the simulation was varied across 1, 2, 3 and 4 times the fly time, that is, $\tau = 10^{-5}$ s, $2 \times 10^{-5}$ s, $3 \times 10^{-5}$ s and $4 \times 10^{-5}$ s, respectively. Thousand runs with independent noises were repeated for each scenario to calculate the probability $p_E$ of Eve's successful guessing of the bit value.

Below, several transient attack/defense situations are demonstrated with various suppression levels of the information leak.

**Scenario-1: No defense: Start at random point in time**

In order to establish a baseline, first the *no-defense* case was simulated, where the driving noises of the bit exchange are started abruptly at random and independent voltage amplitudes at the two ends. The attack is based on the observation of mean-square voltages and currents, see Equations (11) and (12), and it indicates a huge information leak with $p_{E,V}$ and $p_{E,I}$ (the $p_E$ values for voltage and current transients, respectively) around 0.9, see Table I.

We note that an alternative way of extracting information would be by utilizing Equations (9) and (10) for the observed cable reflections due to different terminating resistances. We have not explored this path because it was not necessary, instead we show the very well visible reflections at the multiples of the fly time of $10^{-5}$ s, see Fig. 3.

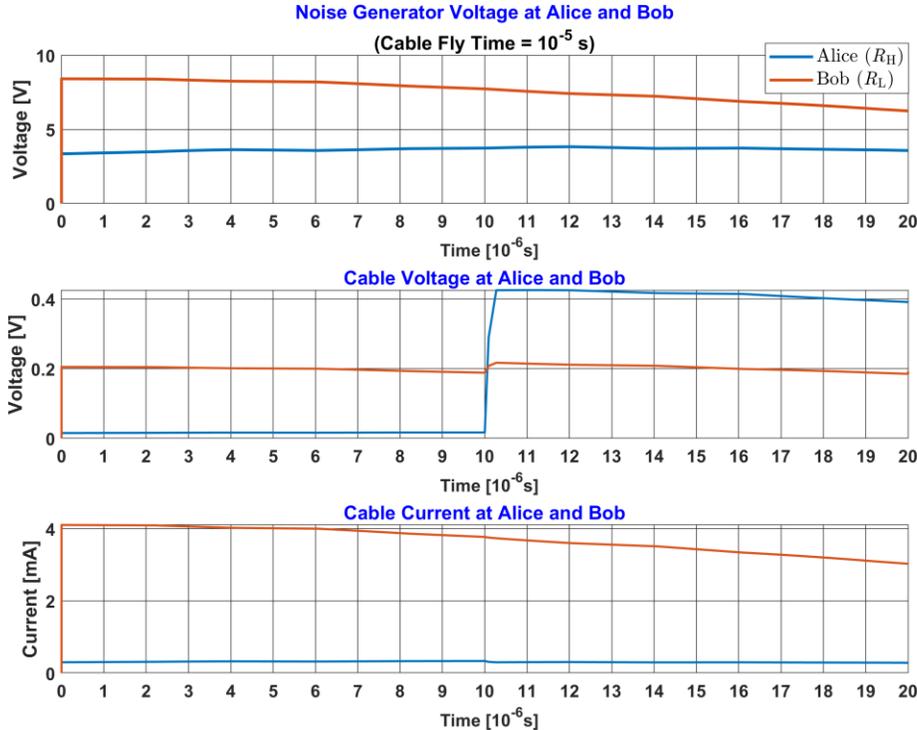

**FIG. 3.** Demonstration of the security vulnerability due to cable reflections when Alice's and Bob's noise voltage generators start from random values. Waveforms at Alice and Bob sides are represented by blue and brown, respectively. The top figure represents the instantaneous noise generator voltages of Alice and Bob, respectively; the middle figure shows the instantaneous cable voltages at Alice and Bob, respectively, and the bottom figure represents the instantaneous cable currents at Alice and Bob, respectively. The simulation was performed over a transient observation time of $\tau = 2 \times 10^{-5}$ s. The parameters of the simulation: $R_H$ = 11 KΩ, $R_L$ = 2 KΩ, $T$ = 7x10$^{15}$ K, $B$ = 5 KHz, $Z_0$ = 50 Ω and $t_f$ = 10$^{-5}$ s. The step functions are due to cable reflections, see Equations (9) and (10), which can yield information about the resistance values of the different cable terminations ($R_H$ and $R_L$, see Fig. 1) that are causing the reflections.

| Case | $\tau$ ($10^{-6}$s) | $p_{E,V}$ | $p_{E,I}$ |
|------|------|------|------|
| A | 10 | 0.890 ± 0.008 | 0.890 ± 0.008 |
| B | 20 | 0.875 ± 0.011 | 0.905 ± 0.009 |
| C | 30 | 0.848 ± 0.015 | 0.918 ± 0.009 |
| D | 40 | 0.848 ± 0.010 | 0.927 ± 0.007 |

**TABLE I.** Demonstration of the mean-square voltage based transient attack, see Equations (11) and (12), with no defense being applied, over four different transient observation time ($\tau$). Case (A), (B), (C) and (D) represent transient observation time of $\tau = 10^{-5}$ s, $2\times10^{-5}$ s, $3\times10^{-5}$ s and $4\times10^{-5}$ s, respectively. Parameters of the simulation: $R_H$ = 11 KΩ, $R_L$ = 2 KΩ, $T$ = 7x10$^{15}$ K, $B$ = 5 KHz, $Z_0$ = 50 Ω and $t_f = 10^{-5}$ s. In all these cases, Eve's probability $p_E$ of correctly guessing the bit is about 0.9 indicating a huge information leak.

**Scenario-2: Partial defense attempt: Starting at zero voltage, but no slope match**

Here Alice and Bob's noise voltages are started from zero voltage, but the starting slopes are not matched. The attack is again based on the observation of mean-square voltages and currents, see Equations (11) and (12). Similarly, to the no-defense situation, the $p_E$ value is around 0.9 (see Table II). Thus, this way of defense does not work either.

| Case | $\tau$ ($10^{-6}$s) | $p_{E,V}$ | $p_{E,I}$ |
|------|------|------|------|
| E | 10 | 0.949 ± 0.007 | 0.949 ± 0.007 |
| F | 20 | 0.909 ± 0.007 | 0.945 ± 0.005 |
| G | 30 | 0.889 ± 0.008 | 0.939 ± 0.012 |
| H | 40 | 0.880 ± 0.007 | 0.938 ± 0.009 |

**TABLE II.** Demonstration of the mean-square voltage based transient attack, see Equations (11) and (12); when the two noise generators start from zero voltage, but with arbitrary slope. The simulation was performed over four different transient observation time ($\tau$). Case (E), (F), (G) and (H) represent transient observation time of $\tau = 10^{-5}$ s, $2\times10^{-5}$ s, $3\times10^{-5}$ s and $4\times10^{-5}$ s, respectively. Parameters of the simulation: $R_H$ = 11 KΩ, $R_L$ = 2 KΩ, $T$ = 7x10$^{15}$ K, $B$ = 5 KHz, $Z_0$ = 50 Ω and $t_f = 10^{-5}$ s. In all these cases, Eve's probability $p_E$ of correctly guessing the bit is about 0.9 indicating a huge information leak.

**Scenario-3: Partial defense: The starting generator voltages and slopes are maintained at the ratio of $m_{HL}$**

Alice and Bob's noise voltages begin from a non-zero value; where their starting noise voltage and starting slope are maintained at the ratio of $m_{HL}$, see Equation (8). The attack is again based on the observation of mean-square voltages and currents, see Equations (11) and (12). Simulations were performed over four different transient observation time of $\tau = 10^{-5}$ s, $2\times10^{-5}$ s, $3\times10^{-5}$ s and $4\times10^{-5}$ s, respectively. If Eve only has up to the first fly time, she does not get sufficient time to crack the KLJN scheme; Eve's probability ($p_{E,V}$ and $p_{E,I}$) to successfully guess the proper key bit is around 0.5 (perfect security). The reduced $p_E$ (see Table III) indicates that this method offers significant defense compared to Scenarios 1 and 2.

| Case | $\tau$ ($10^{-6}$s) | $p_{E,V}$ | $p_{E,I}$ |
|------|------|------|------|
| I | 10 | 0.502 ± 0.012 | 0.502 ± 0.012 |
| J | 20 | 0.763 ± 0.013 | 0.699 ± 0.013 |
| K | 30 | 0.603 ± 0.011 | 0.665 ± 0.011 |
| L | 40 | 0.629 ± 0.013 | 0.627 ± 0.015 |

**TABLE III.** Demonstration of the mean-square voltage based transient attack, see Equations (11) and (12), with Alice and Bob's starting generator voltages and slopes maintained at the ratio of $m_{HL}$, see Equation (8). The simulation was performed over four different transient observation time ($\tau$). Case (I), (J), (K) and (L) represent transient observation time of $\tau = 10^{-5}$ s, $2\times10^{-5}$ s, $3\times10^{-5}$ s and $4\times10^{-5}$ s, respectively. Parameters of the simulation: $R_H$ = 11 KΩ, $R_L$ = 2 KΩ, $T$ = 7x10$^{15}$ K, $B$ = 5 KHz, $Z_0$ = 50 Ω and $t_f = 10^{-5}$ s. The reduced $p_E$ indicates that this method offers significant defense compared to Scenarios 1 and 2.

**Scenario-4: Proposed defense: The starting voltage is zero and the slopes are maintained at the ratio of $m_{HL}$**

The results of the ultimate defense method (see section PROPOSED DEFENSE METHODS above) is shown below. Alice's and Bob's generator noises start from zero voltage, and the starting slope of the two noises are scaled at the ratio of $m_{HL}$, see Equation (8). During most of the transient period, the noise voltages and currents in cable ramp up at very similar rate at two ends of Alice and Bob, not allowing Eve to distinguish.

| Case | $\tau$ ($10^{-6}$s) | $p_{E,V}$ | $p_{E,I}$ |
|------|------|------|------|
| M | 10 | 0.506 ± 0.012 | 0.506 ± 0.012 |
| N | 20 | 0.516 ± 0.013 | 0.509 ± 0.013 |
| O | 30 | 0.528 ± 0.023 | 0.514 ± 0.023 |
| P | 40 | 0.531 ± 0.012 | 0.511 ± 0.010 |

**TABLE IV.** Demonstration of our proposed defense against the mean-square voltage based transient attack; Alice and Bob's noise generators are started from zero voltage and with matching slope at the ratio of $m_{HL}$, see Equation (8). The simulation was performed over four different transient observation time ($\tau$). Case (M), (N), (O) and (P) represent transient observation time of $\tau = 10^{-5}$ s, $2 \times 10^{-5}$ s, $3 \times 10^{-5}$ s and $4 \times 10^{-5}$ s, respectively. Parameters of the simulation: $R_H$ = 11 KΩ, $R_L$ = 2 KΩ, $T$ = 7x10$^{15}$ K, $B$ = 5 KHz, $Z_0$ = 50 Ω and $t_f = 10^{-5}$ s. In all 4 cases, Eve does not have sufficient information to crack the KLJN scheme, as $p_{E,V}$ and $p_{E,I}$ are close to 0.5.

## CONCLUSION

It is demonstrated by cable simulations that the ideal KLJN scheme is vulnerable against the transient attack. The best defense introduced in this paper is the by proper choice of the starting voltage and starting slope of the two noise generators. This method can successfully defend against the transient attack.

## DATA AVAILABILITY

The data that support the findings of this study are available from the corresponding author upon reasonable request.